# Interannual observations and quantification of summertime H$_2$O ice deposition on the Martian CO$_2$ ice south polar cap

Adrian J. Brown[*1], Sylvain Piqueux[2], Timothy N. Titus[3]

[1] SETI Institute, 189 Bernardo Ave, Mountain View, CA 94043, USA
[2] Jet Propulsion Laboratory, California Institute of Technology, 4800 Oak Grove Drive, Pasadena, CA 91109 USA
[3] U.S. Geological Survey, Astrogeology Science Center, Flagstaff, AZ 86001 USA

## Abstract

The spectral signature of water ice was observed on Martian south polar cap in 2004 by the *Observatoire pour l'Mineralogie, l'Eau les Glaces et l'Activite* (OMEGA) (Bibring et al., 2004). Three years later, the OMEGA instrument was used to discover water ice deposited during southern summer on the polar cap (Langevin et al., 2007). However, temporal and spatial variations of these water ice signatures have remained unexplored, and the origins of these water deposits remains an important scientific question. To investigate this question, we have used observations from the Compact Reconnaissance Imaging Spectrometer for Mars (CRISM) instrument on the Mars Reconnaissance Orbiter (MRO) spacecraft of the southern cap during austral summer over four Martian years to search for variations in the amount of water ice.

We report below that for each year we have observed the cap, the magnitude of the H$_2$O ice signature on the southern cap has risen steadily throughout summer, particularly on the west end of the cap. The spatial extent of deposition is in disagreement with the current best simulations of deposition of water ice on the south polar cap (Montmessin et al., 2007).

[*] corresponding author, email: abrown@seti.org

This increase in water ice signatures is most likely caused by deposition of atmospheric H$_2$O ice and a set of unusual conditions makes the quantification of this transport flux using CRISM close to ideal. We calculate a 'minimum apparent' amount of deposition corresponding to a thin H$_2$O ice layer of 0.2mm (with 70% porosity). This amount of H$_2$O ice deposition is 0.6-6% of the total Martian atmospheric water budget. We compare our 'minimal apparent' quantification with previous estimates.

This deposition process may also have implications for the formation and stability of the southern CO$_2$ ice cap, and therefore play a significant role in the climate budget of modern day Mars.

**Highlights**

- We report on the H$_2$O ice depositional cycle on the Martian CO$_2$ ice residual south polar cap

- We use data from the CRISM instrument obtained over the past four Martian summer periods

- Potential models: 1) cold trapping 2) sublimation 3) entrained H$_2$O ice in sublimation flow

- The 'minimal apparent' amount of water ice deposited corresponds to a layer 0.2mm thick.

- This amounts to over 0.6-6% of the total Martian atmospheric water budget.

Corresponding author:

Adrian Brown

SETI Institute

189 Bernardo Ave, Mountain View, CA 94043

ph. 650 810 0223

fax. 650 968 5830

email. abrown@seti.org

Short running title: "Observations of the H$_2$O ice cycle on summertime Martian south pole"

transfer

**1. Introduction**

The Martian north polar ice cap has long been understood as the most important exposed source and sink of water on modern day Mars (Farmer et al., 1976). In contrast, the south seasonal cap was thought to be essentially composed of $CO_2$ ice following surface temperature measurements using the Viking Infrared Thermal Mapper (IRTM) (Kieffer, 1979). In the late 1990s, the Mars Global Surveyor Thermal Emission Spectrometer (TES) also confirmed $CO_2$ ice temperatures in the south pole residual cap (SPRC) during the austral summer (Kieffer et al., 2000) and found "no significant presence" of $H_2O$ ice. Soon after this observation, Nye et al. (2000) published an influential theoretical paper showing that a pure $CO_2$ ice south polar cap would collapse under its own weight, and suggested dirty $H_2O$ ice as the principle ice cap constituent. In 2004, the OMEGA instrument on Mars Express was used for one Earth month to observe the Martian south polar region from $L_s$=335-348 and reported observations of $H_2O$ ice mixed with $CO_2$ ice in the residual south polar cap and polar layered deposits (Bibring et al., 2004). Doute et al. (2006) used OMEGA to examine the water and dust content of the south polar cap during late summer. Langevin et al. (2007) mapped the springtime retreat and summer evolution of the south polar cap and were the first to note the deposition of water ice on the south polar cap.

Further complicating this cap compositional picture, large deposits of $CO_2$ ice have recently been discovered in the subsurface of the south polar layered deposits (Phillips et al., 2011).

Viking IRTM and TES temperature observations show that $CO_2$ ice is the predominating ice in the surface/near surface in the residual cap during summer (Kieffer, 1979; Titus et al., 2008). Mars Global Surveyor Mars Orbiter Camera (MOC) images of the south polar cap were used to discover 'Swiss cheese features' (Malin et al., 2001) which were successfully modeled as a meters-thick layer of $CO_2$ ice underlain by $H_2O$ ice (Byrne and Ingersoll, 2003). Mars Odyssey Thermal Imaging Spectrometer (THEMIS) temperature observations of the south polar cap have been used to infer that $H_2O$ ice becomes exposed at the periphery of the residual cap during late summer (Piqueux et al., 2008), and Titus et al. (2003) used thermal and visual observations, combined with subpixel mixing models, to identify a possible $H_2O$ ice lag at the edge of the gullies and trenches in the SPRC that were exposed as the edge of the retreating seasonal $CO_2$ ice cap moved poleward during summer. However, THEMIS and TES do not have spectral coverage of near-infrared (NIR 1.0-2.5 μm) region of the electromagnetic (EM) spectrum where several $CO_2$ and $H_2O$ ice absorption bands are available to CRISM.

CRISM is very sensitive to water ice deposited on the surface of Mars. For example, CRISM data were recently used to observe an unexpected asymmetric springtime retreat of $CO_2$ ice observed in the north polar cap, potentially due to $H_2O$ ice sourced from the north polar cap outliers (Brown et al., 2012). CRISM has also been used to examine Rayleigh scattering in the Martian atmosphere (Brown, 2014) and to discriminate water and $CO_2$ ice deposited in halos around Swiss-cheese deposits (Becerra et al., 2014). Here we present CRISM observations of the south polar cap over four Martian years that show that surficial $H_2O$ ice reappears each year on a repeatable and cyclic basis on the residual $CO_2$ ice cap throughout the austral summer. The increases in $H_2O$ deposition occur across the entire SPRC but not in dusty or regolith-dominated regions beyond it.

Placing limits on the modern day deposition rate of water ice on the SPRC will allow us to better understand the modern-day stability of the southern polar cap (Byrne, 2009) and perhaps even shed light on the formation processes on the south polar cap (Montmessin et al., 2007). Many simulations of the Martian water cycle have investigated the question of cold-trapped water ice on the exposed $CO_2$ ice (Jakosky, 1983; Haberle and Jakosky, 1990; Houben et al., 1997; Richardson and Wilson, 2002; Montmessin et al., 2004; Montmessin et al., 2007) and we conclude the paper by comparing their predictions with our multiyear CRISM measurements.

## 2. Methods

Our primary means of monitoring the SPRC $H_2O$ ice cycle comes from mosaics and spectra constructed of the CRISM global mapping data during Mars Year (MY) 28-31. The Compact Reconnaissance Imaging Spectrometer for Mars (CRISM) is a visible to near-infrared spectrometer on Mars Reconnaissance

Orbiter (MRO) spacecraft that is sensitive to near infrared (NIR) light from ~0.39 to ~3.9 μm and is operated by the Applied Physics Laboratory at Johns Hopkins University. We used the multispectral (MSP and HSP) TRR3 I/F data that are available from the Planetary Data System (PDS).

In CRISM mapping mode 10x on-instrument binning is employed in the cross-track direction. Consequently the mapping swathes we use are 60 pixels across, covering approximately 10.8 km on the surface (Murchie et al., 2007) with a down and cross track resolution of ~182 m. The length of each swathe is controlled by exposure time and is variable depending on commands sent to MRO.

We produced mosaics of all the CRISM mapping data available for each two-week period (equivalent to the time of an MRO planning cycle and also a useful cadence for investigating seasonal change in the Martian polar regions). Each mosaic is in polar-stereographic projection. Figure 1 shows the residual south polar ice cap region as a mosaic of CRISM images of the SPRC during the summer of MY28.

<insert Figure 1 here>

Water ice can be mapped on the surface using these near infrared mosaics by exploiting the 1.5 μm water ice absorption band. In Figure 2 we display a mosaic map for early and late summer from MY 28, showing the presence of $H_2O$ and $CO_2$ ice. We use a $H_2O$ ice index first used by Langevin et al. (2007) and adjusted for use with CRISM by Brown et al. (2010a). The formula for this index is:

$$H_2O index = 1 - \frac{R(1.500)}{R(1.394)^{0.7} R(1.750)^{0.3}} \qquad (1)$$

Where $R(\lambda)$ indicates the reflectance at the wavelength $\lambda$ in µm. The index is high when the 1.5 µm water ice band is present and low when it is not, and it increases as larger grain water ice is present(Warren, 1982).

<insert Figure 2 here>

## 3. Results

*3.1 Residual ice cap mosaics*

The first line of evidence for $H_2O$ ice deposition is the residual cap ice identification maps in Figure 2. Following previous studies (Brown et al., 2010a; Brown et al., 2012) we use a $H_2O$ index threshold of 0.125 to indicate that surficial $H_2O$ is present. $CO_2$ ice is detected using a band analysis routine described (Brown, 2006) and successfully applied by Brown et al. (Brown et al., 2008b; Brown et al., 2010b). As in previous studies (Brown et al., 2010a; Brown et al., 2012), positive identification of $CO_2$ ice is indicated by a threshold 1.435 µm band depth of 0.16.

The ice identification maps show almost complete coverage of the residual cap in $CO_2$ ice (shown in red) and little to no $H_2O$ ice in the early summer period ($L_s$=310). On the right of Figure 2, we have superposed the CRISM observations in the $L_s$=310-330 time period. These show strips of cyan where CRISM has detected small but significant increases in the 1.5 $H_2O$ µm band. As can be seen from the CRISM mosaics in Figure 2, the effect is confined sharply to the south polar cap, and does not extend beyond the SPRC edge.

<insert Figure 3 here>

*3.2 CRISM Reflectance Spectra*

Our second line of evidence is taken from spectra extracted from a point within these maps. Figure 3 shows a set of individual CRISM MSP spectra (no averaging or binning has been done) prior to $H_2O$ ice deposition (starting at $L_s$=261) and throughout summertime (finishing at $L_s$=337) from 265.5°E, 86.1°S (marked as Point A on Figure 1). The spectra show a marked decrease in the 1.5 µm band that causes a decrease in the shoulder of the $CO_2$ ice absorption band at 1.4 µm. As in previous studies (Brown et al., 2008a; Brown et al., 2010a; Brown et al., 2012) we attribute this change in band depth to the presence of increasing amounts of $H_2O$ ice through the summer season.

<insert Figure 4 here>

*3.3 Point observations at Point A and B*

Our third line of evidence comes from mapping observations taken by the CRISM instrument for MY28-31. In order to establish that this process is cyclic, we have extracted individual spectra from regions close to Point A and have plotted the $H_2O$ ice and $CO_2$ ice index and also to the 1.2 µm albedo for these spectra in Figure 4a-c. Coverage in MY 29-31 is not quite as comprehensive, however in Figure 4d-f we show the same data for Point B, which is on the other side of the SPRC from Point A. Being closer to the south pole, the composition of the Point B region is dominated by $CO_2$ ice for the duration of summer, and smaller amounts of $H_2O$ ice signature appear at this location. However, as for Point A, a definite $H_2O$ ice cycle can be observed throughout summertime, indicated by a $H_2O$ ice index that starts relatively low and increases steadily throughout summertime. Thus, we were able to establish that the $H_2O$ index behaves in a similar manner (i.e. it increases) throughout the summer across the entire SPRC and for all Martian years for which we have data.

<insert Figure 5 here>

*3.4 Radiative Transfer Model and Quantification of Water Ice Deposited*

In order to quantify the amount of $H_2O$ ice deposited on the south polar cap, we developed a radiative transfer model to reproduce the broad characteristics of a number of CRISM MSP spectra taken at Point A (see Figure 1). The CRISM spectra taken at $L_s$=337 (in late summer, after $H_2O$ ice deposition) in MY28 and radiative transfer model spectra are shown in Figure 5.

We carried out the radiative transfer modeling using the model proposed by Shkuratov et al.(1999) which is a simplified 1-dimensional model that allows us to interpret the effect of $H_2O$ ice as a contaminant in a $CO_2$ ice snowpack. As in past research (Brown et al., 2010a), we used a three component model (optical constants of $CO_2$ ice (Hansen, 2005), $H_2O$ ice (Warren, 1984) and palagonite (Roush et al., 1991) as a Martian dust simulant) and attempted to fit the overall albedo and key band strengths of $CO_2$ ice and $H_2O$ ice in the spectrum.

We carried out two modeling approaches in order attempt to place physical bounds on the amount of water ice deposited on the south polar cap. The two models adopted are:

1.) 'minimum apparent' model, and

2.) 'extrapolated' model

The 'minimum apparent' model attempts to find the smallest amount of water ice that could explain the observed water ice band, and as discussed in previous work (Brown et al., 2010a; Brown et al., 2012) this is the most robust way to interpret the observed CRISM spectra. The 'extrapolated' model makes additional assumptions in an effort to provide a reasonable best approximation to the depth of the $H_2O$ ice deposits.

'Minimum apparent' model. To carry out the 'minimum apparent' interpretation, we used a newly developed iterative band-fitting routine based on previous band fitting algorithms (Brown, 2006) to generate a three-component mixture of $CO_2$, $H_2O$ ice and dust that matched three components:

1.) the NIR albedo,

2.) the $CO_2$ ice band depths at 1.435 μm, and

3.) the doublet near 2.2 μm and $H_2O$ ice band at 1.5 μm.

For the late summer spectrum in Figure 5, we found that the best spectral match was obtained for $CO_2$ ice grain size of ~4.25mm, ($C_{CO2}$=85.7% by volume) $H_2O$ ice grain size of ~ 0.2mm ($C_{H2O}$=5.1%) and palagonite of ~0.35 mm ($C_{dust}$=9.2% by volume). We used a porosity (q-factor) of $q$=0.3, corresponding to 70% pores and 30% ice/dust. These results are summarized in Table 1.

| Component | Reference | Best fit grain size (microns) | Best fit concentration |
|---|---|---|---|
| $CO_2$ ice | Hansen (2005) | 4252.7 | 0.857 |
| $H_2O$ ice | Warren (1984) | 200.1 | 0.051 |
| Palagonite (soil) | Roush et al. (1991) | 354.9 | 0.092 |

Table 1 – Details of the best fit parameters for the CRISM spectrum in Figure 5. The fit was applied over the spectral range from 1.02-2.5 microns. In this range, using wavelengths from the CRISM MSP range, there are 42 bands. Over this range, the average absolute fitting error per band is 0.01659 for this best fit. The porosity was constrained to be 0.3 (30% ice, 70% vacant).

In order to estimate the 'minimum apparent' amount of water ice deposited, we make the simplest assumption that the thickness of the water ice layer is a minimum of the derived grain size diameter ($D_{H2O}$=0.2mm or $2 \times 10^{-7}$km) from the radiative transfer calculation. The physical justification for this is that the water ice must be optically detectable, and therefore at least one optical pathlength should be available for the passage of vertically propagating photons. The true situation will be far more complex. We make the obviously simplified assumptions that:

1.) the residual SPRC has an approximate area of $A_{cap}$=2x10$^5$ sq. km (Brown et al., 2010a),

2.) the SPRC is uniformly covered in summertime by a water ice layer, and

3.) that the $H_2O$ ice is distributed in a 'checkerboard' fashion and occupies $C_{H2O}.q$ = 0.05 * 0.3 = 0.015 or 1.5% of this area.

With these three assumptions, the minimal amount of water ice observed would be determined by the following linear relationship:

$$volume_{H_2O} = A_{cap}C_{H_2O}D_{H_2O}q \qquad\qquad (2)$$

therefore $volume_{H2O\text{-}minimal\ apparent}$=2x10$^5$*0.05*2x10$^{-7}$*0.3=6x10$^{-4}$ km$^3$.

We regard this as a 'minimum apparent' estimate because it is likely that the deposit could be as thick as 10 grain diameters (10.$D_{H2O}$).

<u>Extrapolated estimate</u>. The 'extrapolated' approach assumes that the water ice observed by CRISM is the top layer of a deeper deposit that extends ten grain diameters, which is a reasonable situation for this type of deposit. In this case, using:

$$volume_{H_2O-extrapolated} = A_{cap}C_{H_2O}10D_{H_2O}q \qquad\qquad (3)$$

we find $volume_{H2O\text{-}extrapolated}$=2x10$^5$*0.15*6x10$^{-8}$*0.3=6x10$^{-3}$ km$^3$.

These rather firm sounding numbers must be balanced with the fact that we have assumed that the entire SPRC is covered by the same amount of water ice, whereas we know from observations that the western part of the polar cap is covered in more ice than the eastern part of the cap (Figure 2).

Nevertheless, assuming a Martian atmospheric $H_2O$ ice budget of ~0.1 km$^3$ (Christensen, 2006), the amount of water ice participating in this process will make up as much as 0.6% (at minimum, for 0.2mm thick ice deposit) to 6% (at maximum, for 2mm thick ice deposit) of the atmospheric Martian water budget. Note that we are not stating that 0.6-6% of the atmospheric water ice is being deposited on the cap, because we cannot be sure of the immediate source of the process. However, if the water ice is to be completely sourced from the atmosphere, then it would make up 0.6-6% of the current Martian atmospheric water budget.

## 4. Discussion

*4.1 Comparison with previous hydrological models*

| Reference | Annual Water loss rate to south polar cap in grams | Total Annual Water loss rate to south polar cap in microns | Transport flux to southern polar cap |
|---|---|---|---|
| Jakosky and Farmer (1982) | $4 \times 10^{14}$ g (maximum) | | |
| Jakosky (1983) p.37 | $1.4$-$4.1 \times 10^{14}$ g | | 20-40% |
| Haberle and Jakosky (1990) (their Table 2) | | 100-800 (lost by north cap) | 12-25% |
| Houben (1997) p.9078 | $2.5 \times 10^{14}$ g or 0.25GT | 500 | 12-25% |
| Richardson and Wilson (2002) (their "VS Study") | $1.4 \times 10^{14}$ g | | 7-14% |
| Montmessin et al. (2007) p.8 | | 400 | 9.6-20% |
| This study | $0.6$-$6 \times 10^{13}$ (min app) | | 0.6 (min app) |

| | 1.2-12x10$^{13}$ $_{(extrap)}$ | | 6 $_{(extrap)}$% |
|---|---|---|---|

Table 2 – Comparison of GCM model water loss to south polar cap amounts with the results of this study. For the purposed of calculating the transport flux, the estimated total atmospheric water content is estimated at 1-2 x10$^{15}$ g.

We compare this estimate with previous Martian hydrological cycle models in Table 2. Jakosky and Farmer (1982) made the first upper estimate of the water that might be transported to the south polar cap by measuring the water vapor above the north polar cap using the MAWD instrument on the Viking orbiters. Jakosky (1983) developed a simplified circulation model that reproduced observations of the MAWD instrument and included a loss of water ice to the south polar cap and a regolith sink. Haberle and Jakosky (1990) used a simulation of the Martian north pole in the light of MAWD measurements to suggest that regolith was necessary in keeping the north polar cap stable. They provided an estimate of 0.1-0.8 mm of loss of water ice from the north polar water ice cap, which they used to put upper bounds on the amount deposited on the south polar cap. Houben et al. (1997) developed a simplified 3D climate model of the Martian water cycle which included transport between atmosphere and regolith. Richardson and Wilson (2002) reported the first use of a Martian GCM with a water ice cycle, with water ice treated as a trace component, and Montmessin et al. (2004) carried out a similar study including water ice clouds (with varying size distributions) with the LMD GCM.

Our estimates are smaller than the estimates of all GCM models, which may be due to the conservative approach taken to interpretation of our spectra. We consider this a first order estimate that will be refined as future observations are made of the summer south polar cap.

Montmessin et al. (2007) carried out a similar GCM-based simulation of the amount of water ice deposited on the south polar cap on current-day Mars and compared this to a model of water ice deposition during reversed perihelion. They suggested the water ice at the base of the SPRC was emplaced during reversed perihelion conditions.

Montmessin et al. (2007) used a symmetric model of the south polar cap and predicted that water ice deposition would correlate with latitude – hence greatest deposition is at the pole and smaller amounts of $H_2O$ ice would be deposited at the edge of the cap. This is contrary to the CRISM observations reported here of deposition favoring the warmer western part of the cap, a trend which is also apparent in the OMEGA $H_2O$ ice maps presented by Langevin et al. (2007) (their Figure 18-19).

These observational versus simulation discrepancies indicate to us that future mesoscale modeling of this season is required. At the very least, higher resolution hydrological simulations with a realistic cap orientation should be used to help interpret the observed western depositional pattern on the south polar cap under modern Martian conditions.

*4.2 Interannual differences*

During MY28 there was a global dust storm (James et al., 2010) that induced a decrease in the albedo of the polar ice and subdued the $H_2O$ and $CO_2$ ice signatures across the SPRC. This explains the relatively low MY28 index values seen in Figure 4. We have looked for evidence of larger amounts of $H_2O$ ice being deposited each year but the evidence is inconclusive thus far.

*4.3 Seasonal changes*

The data presented in Figure 2 suggest that the depositional process is gradual and lasts all summer long, rather than being due to isolated or singular depositional events each summer. Close to Point A, $H_2O$ index values start around 0.02 near $L_s$=275 and climb to greater than 0.275 at the end of summer for MY29-31. At Point B, which is nearer to the pole, the $H_2O$ index values start around 0.02 near $L_s$=275 and climb to around 0.2 for MY29-31.

*4.4 Deposition or Exposure of $H_2O$ ice?*

It should be noted that the spectra in Figure 3 show $CO_2$ ice and weak $H_2O$ ice signatures mixed together in the same pixel. The mixing might be 'checkerboard' linear mixing or intimate mixing where $CO_2$ ice and $H_2O$ ice grains are encountered by a single photon traversing the Martian snowpack. Therefore, it is not possible for us to tell definitively whether the observed $H_2O$ ice cycle is a process of:

1.) **deposition** of $H_2O$ ice on top of the SPRC (cold trapping)

2.) **sublimation** of $CO_2$ ice, revealing stratigraphically older $H_2O$ ice mixed within the $CO_2$ snowpack.

3.) **atmospheric condensation** of $H_2O$ ice particles within a sublimation flow above the $CO_2$ snowpack.

Option 1 (**deposition**) is our favored explanation for the observations reported here. Thermally thick (e.g. decimeter to meter thick) water ice units have been observed at the periphery and immediate vicinity of the SPRC during the summer following the sublimation of the last seasonal $CO_2$ ice (Titus et al., 2003; Piqueux et al., 2008) and the spectral signature of water ice was also observed by Bibring et al., (2004), on the SPRC mixed with $CO_2$ ice. Titus (2005) pointed out that it is also possible that $H_2O$ ice might be deposited on top of the $CO_2$ ice cap during summertime, thus obscuring a large region of the $CO_2$ ice on the cap. Potential support for this interpretation is the observation that the water ice index is higher in the western regions of the cap, close to where $H_2O$ ice is exposed in the sides of gullies (Titus, 2005).

Option 2 (**sublimation**) is considered less favorable since the weight of GCM modeling suggests that conditions are right for cold trapping of water vapor at this time. The truth may lie between these extremes – water ice may be exposed by $CO_2$ ice sublimation, and then transported to nearby cold trap locations (e.g. tops of 'Swiss-cheese' mesas) where it obscures the underlying $CO_2$ ice. This process may play a role in the burial of large amounts of $CO_2$ ice, which has been a key scientific question arising from the recent SHARAD radar sounder findings of large amounts of $CO_2$ ice beneath the polar layered deposits (PLD) (Phillips et al., 2011).

Option 3 (**atmospheric condensation**) suggests that CRISM is collecting observations of a sublimation flow from the $CO_2$ polar cap during the height of summer, and that small $H_2O$ crystals are forming within the sublimation flow as it evaporates off the ice pack. We consider this option less likely because of the increasing strength of the $H_2O$ ice signature right through to fall – even as conditions begin to cool in late summer.

*4.5 Atmospheric effects*

For this study we have made no effort to remove atmospheric effects from the CRISM data. As mentioned above, a global dust storm was present in MY28 which affected the absorption band depths (Vincendon et al., 2008), however as can be seen in Figure 4, the $H_2O$ cycle continued to operate in a similar manner to the following three Martian years. In addition, the water ice signatures appear only on the $CO_2$ ice cap (unlike the behavior of a cloud). We therefore infer that the $H_2O$ ice cycle is present on the surface, and not due to an atmospheric event (Langevin et al., 2007). If the cycle is a depositional effect, (controlled by on-cap winds), stronger winds may play a role in the SPRC $H_2O$ ice cycle, and this question invites future mesoscale climate modeling of the south polar region.

**5. Conclusions**

The source for the water ice reported here is unclear and we hope this study initiates new avenues of research for the Martian community. These findings impose an important constraint upon models of Martian water ice dynamics while opening a new front in the battle to understand the impoverished but vital Martian water ice cycle.

*5.1. Stability of the SPRC*

The operation of the $H_2O$ ice cycle we have reported may have implications for the stability of the thin veneer of $CO_2$ ice that covers the residual south polar cap. Jakosky and Haberle (1990) suggested that the current $CO_2$ ice south polar residual cap is unstable and could 'flip' quickly to being covered by $H_2O$ ice if water ice from the north polar cap makes its way to the southern cap, which acts as a cold trap in their model. As part of this cycle, and to explain observations of atmospheric $H_2O$ observed above the south pole in 1969 (Barker et al., 1970), these authors suggested that the entire perennial $CO_2$ cap may have disappeared in the summer of 1969 (MY 8) and started to recondense shortly after. However, countering this suggestion, Thomas et al. (2005) used stratigraphic interpretations to date the oldest perennial cap unit as being ~100-150 MY old, excluding a complete removal of the cap in 1969.

Richardson and Wilson (2002) have also determined using GCM modeling that a $H_2O$ ice south polar cap not covered by $CO_2$ ice is unstable under current Martian conditions and would disappear very quickly to the north polar cap.

The physics behind the energy balance of the polar ice cap is clear – in the near infrared (e.g. 1-4 μm) the presence of the $H_2O$ ice layer we have reported here will lead to larger amounts of warming and degradation of the thin $CO_2$ ice veneer of the SPRC. A small amount of $H_2O$ ice will absorb more sunlight in the 1-4μm region (see Figure 3 showing decreasing near infrared albedo of the cap as $H_2O$ index increases), causing more heating of the ice, and therefore more sublimation during the hottest part of the Martian orbital cycle.

*5.2. $CO_2$-$H_2O$ cycle is shown to be steady over 4 Mars years*

It should be pointed out that as seen in Figure 4, for each Martian year observed, the $CO_2$ residual cap has been 're-coated' during winter with $CO_2$ ice. This may have taken place by direct condensation or snowfall during the austral winter (Forget et al., 1998; Hayne et al., 2012; Hu et al., 2012; Hayne et al., 2013). We find 'relatively pure' $CO_2$ ice at the start of each austral summer in our CRISM mosaics (Figure 2). This shows that this 're-coating' process is cyclic and at least stable on observable time scales (Haberle and Jakosky, 1990).

The fact that 'relatively pure' seasonal $CO_2$ ice is present all over the cap indicates that the $H_2O$ ice we see deposited each summertime is entirely interned, and presumably becomes a permanent thin layer within the southern $CO_2$ ice cap. This process may have played a role in forming the south polar layered deposits (SPLD) if they formed at a time when the SPRC was larger.

### 5.3: Varying emissivity of the South Perennial Cap

The albedo and emissivity of the south cap are two critically important parameters determining its stability in the current climate (Wood and Paige, 1992; Blackburn et al., 2009; Guo et al., 2010). A stable cap (able to survive the short but relatively warm southern Martian summers) is difficult to model and previous studies have used various combinations of $CO_2$ ice/dust albedo and emissivities (James and North, 1982; Warren et al., 1990; Hansen, 1999; Doute et al., 2006; Bonev et al., 2008; Kahre and Haberle, 2010; Pilorget et al., 2011; Pommerol et al., 2011; Kieffer, 2013). However, we have shown that the optical properties of the cap transition from "covered by $CO_2$ ice" to "$CO_2$ and $H_2O$ ice mixture" in the course of the Martian austral summer, and future models of the cap albedo and emissivity can now take this variability into account.

### 5.4. The southern cap as a sink for the Martian water cycle

The Martian water cycle is crucial to the understanding of geodynamics of the atmosphere, surface and sub-surface (Clifford, 1993) of the planet. The results of this study show there is an increase in the $H_2O$ ice signature on the south polar residual cap throughout summer, and is distributed predominantly on the western side of the cap. This runs contrary to the results of Montmessin et al. (Montmessin et al., 2007) who modeled a pole-symmetric cap and found that more water ice deposited in the pole, and less deposited on the periphery.

The source of the $H_2O$ ice is uncertain at this stage, but we have proposed three non-exclusive possibilities:

1.) Non-local origin (e.g. transport from the north polar cap)

2.) Local aeolian origin (e.g. exposed water ice around the edges of the cap being blown into the interior and exposed water ice in 'Swiss-cheese' moats (Titus, 2005)).

3.) Local solid-state origin (e.g. sublimation of $H_2O$ ice around the cap periphery or regolith and recondensation over the cold cap (the so-called 'Houben process' in springtime in the north pole (Houben et al., 1997; Brown et al., 2012)).

These three possibilities may all be part of the eventual explanation for this intriguing, widespread and repeatable Martian polar phenomenon.

**Acknowledgements**

We would like to thank the entire CRISM Team, particularly the Science Operations team at JHU APL. We also thank Ted Roush for supplying the palagonite optical constants and Franck Marchis and Peter Jenniskens and referees Yves Langevin and Francois Forget for helpful comments on the manuscript. This investigation was partially funded by NASA Mars Data Analysis Program Grants NNX11AN41G and NNX13AJ73G administered by Mitch Schulte. Piqueux' contribution was carried out at the Jet Propulsion Laboratory, under a contract with the National Aeronautics and Space Administration.

# FIGURES AND CAPTIONS

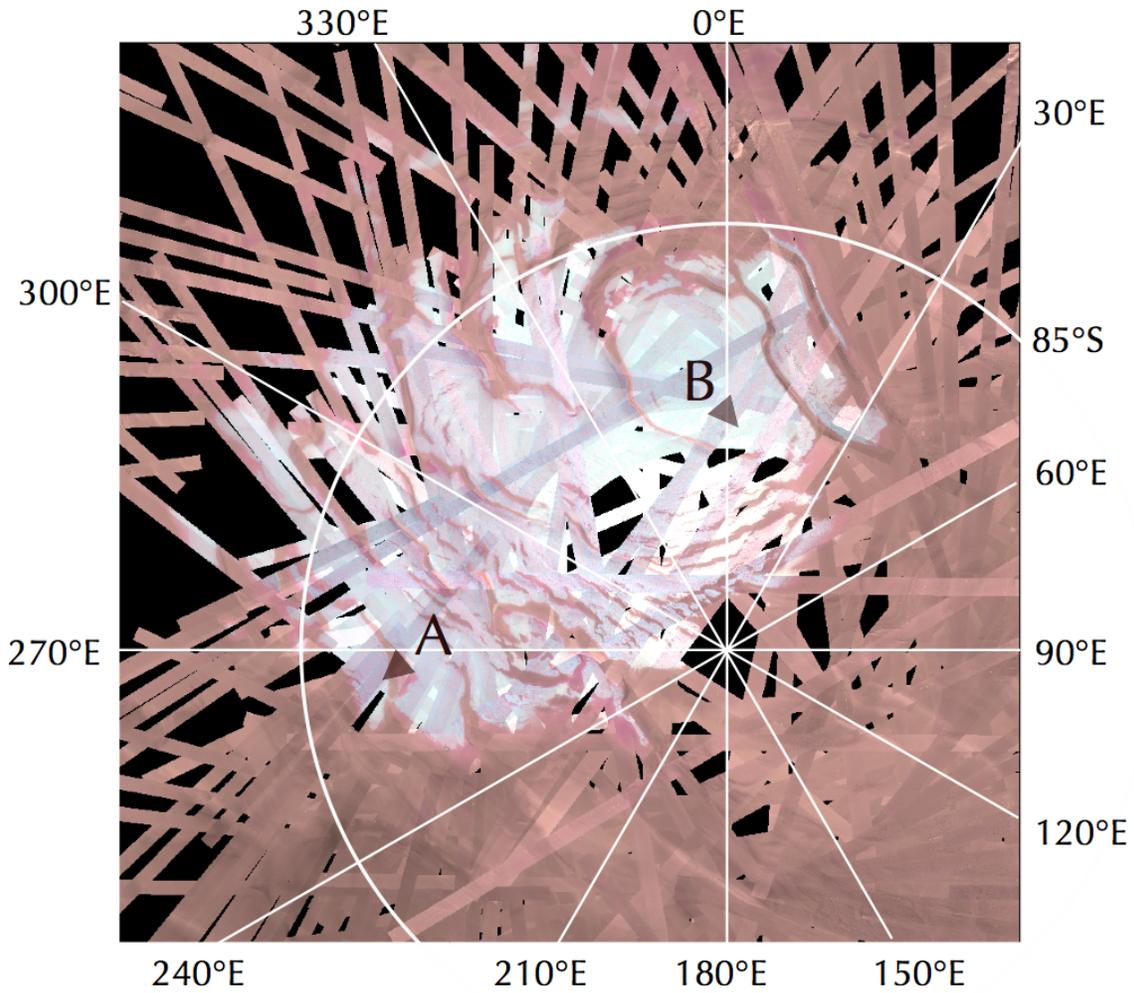

Figure 1 – CRISM mosaic of south polar residual cap (SPRC) in Mars Year 28, compiled during aerocentric longitude $L_s$=304-319 (mid summer) using three CRISM L channel bands (Red: 1.467, Green: 1.427 and Blue: 1.276 m). The locations of Point A (265.5°E, 86.1°S) and Point B (1.0°E, 87.0°S) are shown.

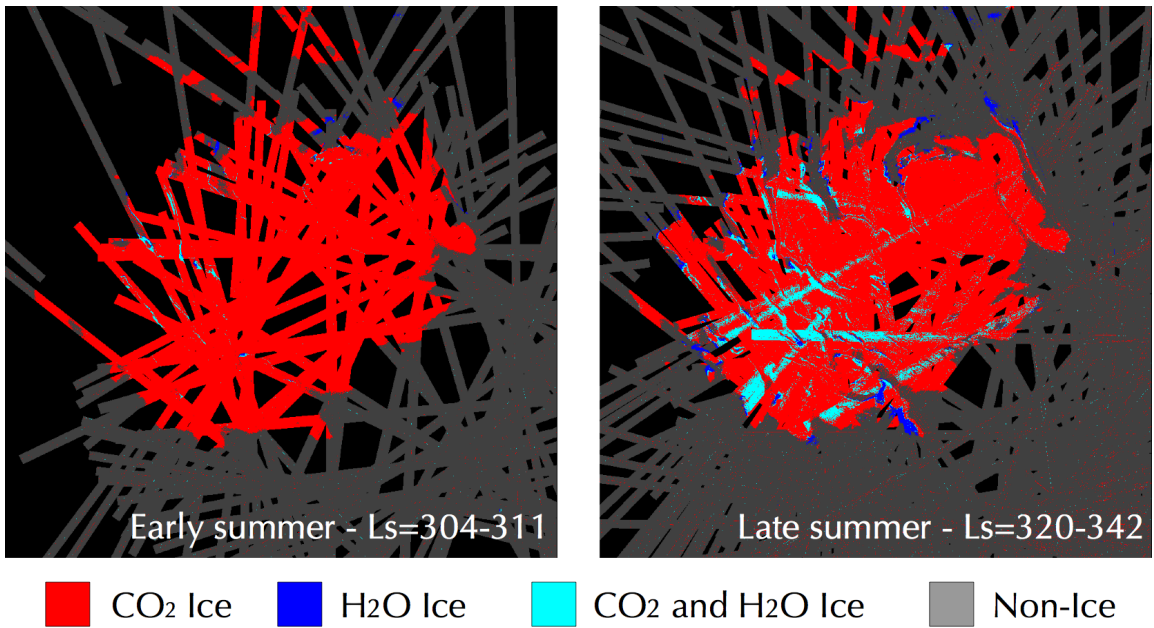

Figure 2 – Martian Year 28 southern summer ice identification mosaics. On left is the mosaic containing images from $L_s$=304-311. Note almost complete coverage by $CO_2$ ice (in red). On right is the mosaic constructed images spanning $L_s$=320-342. Note appearance of $H_2O$ ice mixed with $CO_2$ ice in most recent images.

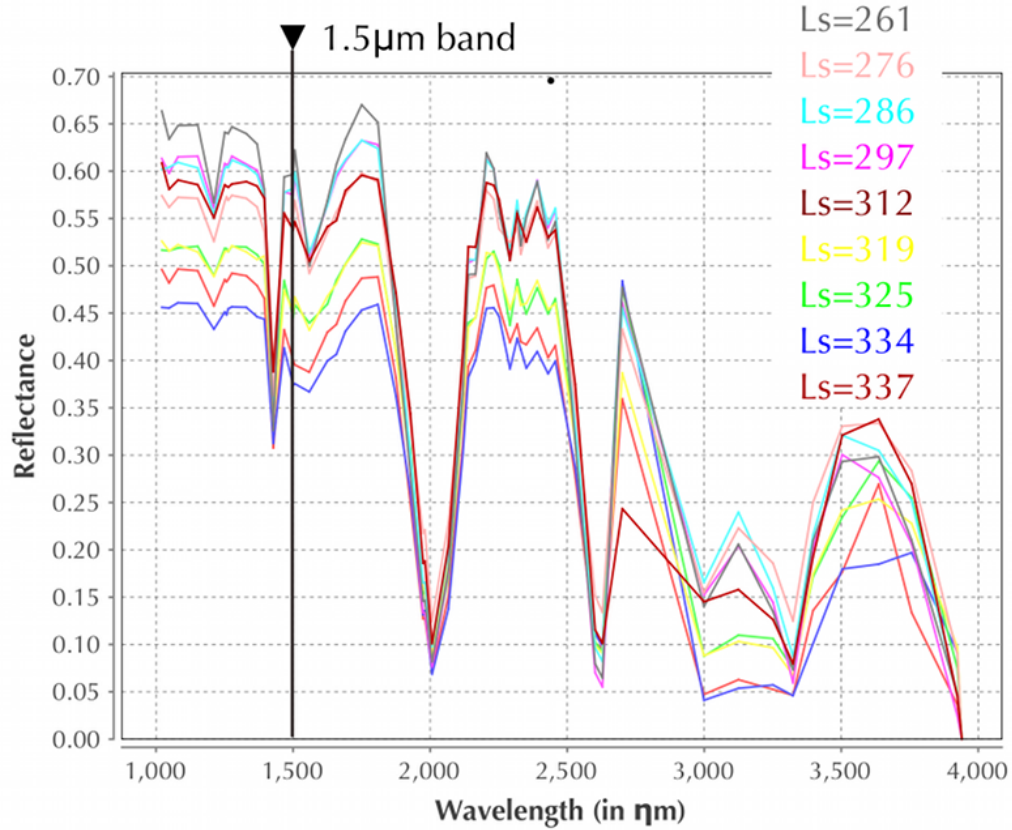

Figure 3 – CRISM summertime MSP spectra (from Point A in Figure 1) showing increase in $H_2O$ ice absorption band for Mars Year 28 in late summer (pixels are ~180m across). The spectra were all taken close to Point A (265.5°E, 86.1°S; see Figure 1). Note overall decreasing albedo and increase in strength of H2O absorption band at 1.5 µm throughout summertime.

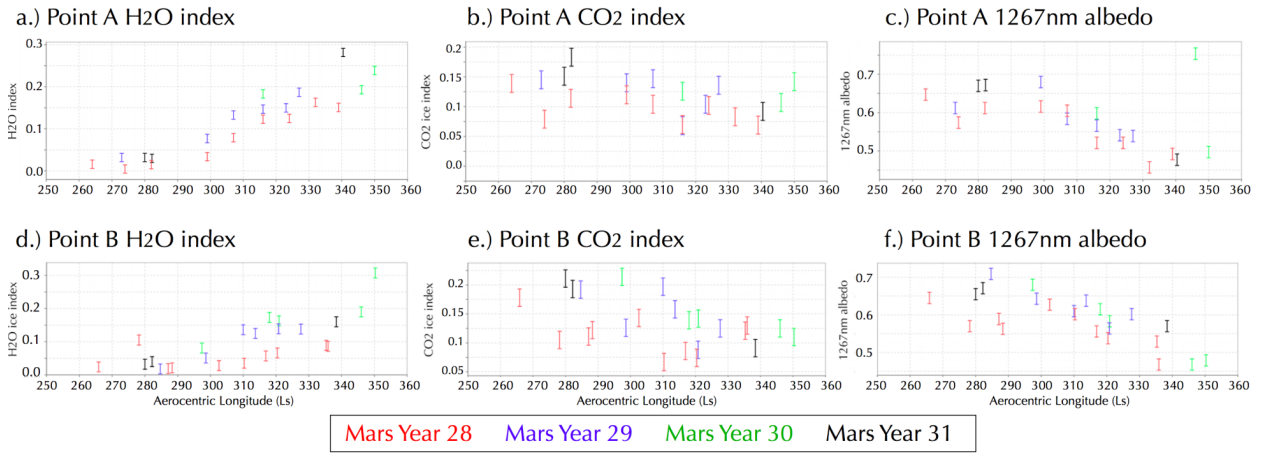

Figure 4a – CRISM $H_2O$ ice index taken from points close to Point A from MY 28-31 during $L_s$=275-360 (austral summer). Note the increasing $H_2O$ index during austral summer (from $L_s$=275-360) across all Mars years where data is available.

Figure 4b-c – CRISM $CO_2$ ice index and CRISM 1.267 µm albedo at Point A during MY 28-31 during austral summer. Figure 4d-f – CRISM $H_2O$ index, $CO_2$ index and 1.267 µm albedo for MY28-31 at Point B, located at (1°E, 87.0°S – see Figure 1).

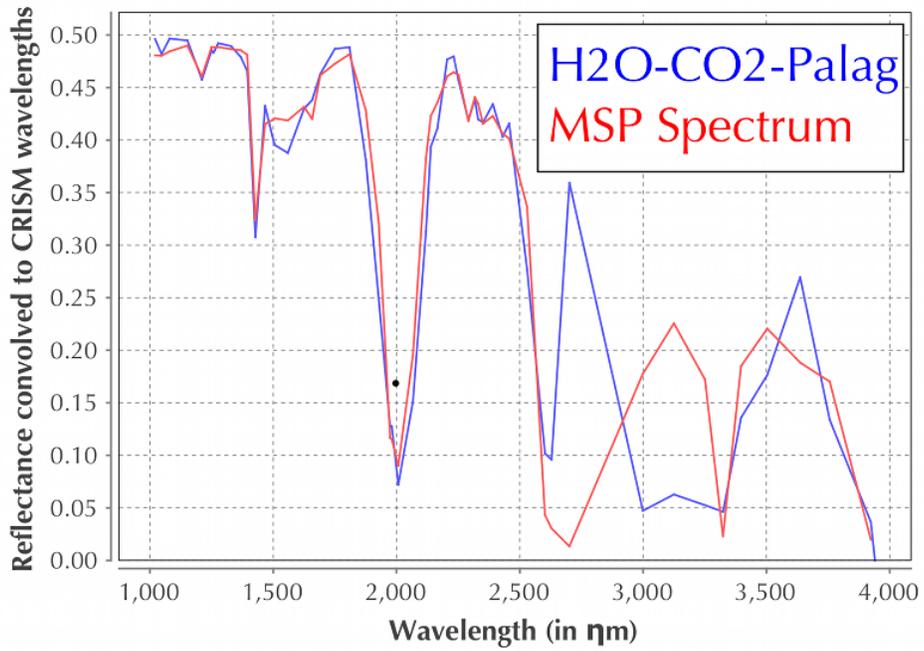

Figure 5 – Shkuratov reflectance models of $CO_2$, $H_2O$ ice and palagonite mixture compared to CRISM L channel MSP spectra. See Table 1 for details of the fit. The CRISM spectra is sourced from Point A (see Figure 1) and was acquired in image MSP 86FC_01 on MY28 $L_s$=337.0 (26 Oct 2007).

## SUPPLEMENTARY INFORMATION

In the course of this investigation, we used CRISM I/F data that had been processed by the CRISM team at JHU/APL to the TRR3 release level. Table A.1 shows the CRISM multispectral summer images that were taken of the Martian south pole from the start of the MRO mission to 27 June 2013.

| | MRO Planning Cycle | DOY (2007-2013) | (MY) and L$_s$ Range | S channel observations | | L channel observations | |
|---|---|---|---|---|---|---|---|
| **Mars Year 28** | 18 | (07)172-181 | 261.5-267.1 | 234 | | 232 | |
| | 19 | 188-198 | 271.5-277.8 | 346 | | 346 | |
| | 20 | 199-212 | 278.4-286.5 | 199 | | 200 | |
| | 21 | 213-219 | 287.1-290.8 | 123 | | 125 | |
| | 22 | 227-240 | 295.6-303.4 | 119 | | 118 | |
| | 23 | 241-254 | 304.0-311.6 | 146 | | 146 | |
| | 24 | 255-268 | 312.2-319.7 | 210 | | 210 | |
| | 25 | 269-282 | 320.2-327.5 | 180 | | 180 | |
| | 26 | 283-296 | 328.1-335.2 | 172 | | 173 | |
| | 27 | 297-309 | 335.7-342.1 | 152 | | 150 | |
| | **Total** | | | **1881** | | **1880** | |
| **Mars Year 29** | 28 | 350-352 | (29) 3.1-4.1 | 26 | | 26 | |
| | 29 | 353-001 | 4.6-10.9 | 183 | | 183 | |
| | 30 | (08)002-012 | 11.4-16.2 | 60 | | 60 | |
| | 31 | 016-026 | 18.1-22.8 | 35 | | 35 | |
| | 58 | (09)113-125 | 252.4-260.0 | 43 | | 43 | |
| | 59 | 127-138 | 260.9-268.4 | 42 | | 42 | |
| | 60 | 140-153 | 269.3-277.5 | 59 | | 59 | |
| | 61 | 154-167 | 277.9-286.3 | 27 | | 26 | |
| | 62 | 168-179 | 286.7-293.5 | 28 | | 28 | |
| | 63 | 184-192 | 296.4-301.6 | 65 | | 66 | |
| | 64 | 196-209 | 303.7-311.6 | 117 | | 116 | |
| | 65 | 210-223 | 312.0-319.7 | 39 | | 39 | |
| | 66 | 224-237 | 319.8-327.6 | 174 | | 174 | |
| | 67 | 238-238 | 327.6-327.7 | 7 | | 7 | |
| | 68 | 350-363 | 24.5-30.7 | 71 | | 71 | |
| | 69 | 364-(10)9 | 30.7-35.6 | 49 | | 49 | |
| | 70 | 013-013 | 37.2-37.3 | 3 | | 3 | |
| | **Total** | | | **1028** | | **1027** | |
| | | | | **MSP** | **HSP** | **MSP** | **HSP** |
| **Mars Year 30** | 91 | (11)82-95 | 259.8-268.3 | 32 | | 31 | |
| | 92 | 99-100 | 270.8-271.2 | 2 | | 2 | |
| | 93 | 114-116 | 279.8-281.6 | 12 | | 12 | |
| | 94 | 128 | 288.7 | 1 | | 1 | |
| | 95 | 142-145 | 297.0-298.8 | 1 | 8 | 1 | 8 |
| | 96 | 170-179 | 313.6-319.2 | 34 | 3 | 34 | 3 |
| | 97 | 180-183 | 319.5-321.3 | 16 | | 16 | |
| | 98 | 226-235 | 344.4-349.4 | 81 | | 81 | |
| | 99 | 236-239 | 349.7-351.5 | 20 | | 20 | |
| | 101 | 282-290 | (31)12.5-16.7 | 4 | | 4 | |
| | 102 | 293 | 17.9 | 1 | | 1 | |
| | **Total** | | | **203** | **11** | **202** | **11** |
| **Mars Year 31** | 123 | (13)013-022 | 244.0-250.2 | 119 | | 119 | |
| | 124 | 023-026 | 250.2-252.6 | 41 | | 41 | |
| | 125 | 056 | 271.2 | 1 | | 1 | |
| | 126 | 065-075 | 277.1-283.5 | 69 | 41 | 69 | 41 |
| | 127 | 167-176 | 336.4-341.6 | 32 | | 32 | |
| | **Total** | | | **262** | **41** | **262** | **41** |

**Table A.1** - CRISM observations of Mars south pole in the near-summertime from MY 28, L$_s$=260 to MY 31 L$_s$=342. L$_s$ = Deg of solar longitude. Southern summer starts at L$_s$ = 270 and ends at L$_s$ = 360.

Locations and observation dates of the spectra shown in Figure 3:

| Ls | MSP ID | MY | MRO Cycle | Earth Day |
|---|---|---|---|---|
| 261.45 | 6508_01 | 28 | 18 | 2007_172 (21 Jun 07) |
| 275.64 | 69F5_05 | 28 | 19 | 2007_194 (13 Jul 07) |
| 285.85 | 6D6E_03 | 28 | 20 | 2007_210 (29 Jul 07) |
| 296.49 | 7256_03 | 28 | 22 | 2007_228 (16 Aug 07) |
| 311.67 | 79AA_01 | 28 | 23 | 2007_253 (10 Sep 07) |
| 318.79 | 7D81_01 | 28 | 24 | 2007_266 (23 Sep 07) |
| 324.60 | 8040_01 | 28 | 25 | 2007_276 (3 Oct 07) |
| 333.67 | 850B_01 | 28 | 26 | 2007_293 (20 Oct 07) |
| 337.0 | 86FC_01 | 28 | 27 | 2007_299 (26 Oct 07) |

Table A.2 Locations and observation dates of the spectra shown in Figure 3.

**Error bars on Figure 4a-f**

Error bars are plotted as constant for the plots of 1267nm, $CO_2$ and $H_2O$ ice index (Fig. 4a-f) as +/-0.015, based on the CRISM signal to noise (~100, based on data in Murchie et al. (Murchie et al., 2007)) for the 1.2-1.5 $\mu$m region.